\documentclass[fleqn,twoside]{article}
\usepackage{espcrc2}

\usepackage{graphics}
 
\begin{document}

\title{Splitting of the superconducting transition in the two weakly coupled
       2D $XY$ models}
\author{Kateryna Medvedyeva, Beom Jun Kim, Petter Minnhagen\\
  Department of Theoreticla Physics,  Ume{\aa} University, 901 87
  Ume{\aa}, Sweden}
%\preprint{\today}

\begin{abstract}
The frequency $\omega$ and temperature $T$ dependent complex conductivity $\sigma$
of two weakly coupled 2D $XY$ models subject to the RSJ dynamics is studied through
computer simulations.  A double dissipation-peak structure in
$\mbox{Re}[\omega\sigma]$ is found as a function of $T$ for
a fixed frequency.  The characteristics of this double-peak structure,
as well as its frequency dependence, is
investigated with respect to the difference in the critical temperatures of the
two $XY$ models, originating from their different coupling strengths. 
The similarity with the experimental data in Festin {\it et al.}
[Physica C {\bf 369}, 295 (2002)] for a thin YBCO
film is pointed out and some possible implications are suggested.
\end{abstract}

\maketitle

The thermally excited two-dimensional (2D) vortex fluctuations drive the phase
transition between the superconducting and normal state in many 2D systems.
Such transition is of the Kosterlitz-Thouless (KT) type: below the
KT transition temperature $T_{c}$ vortices are bound in pairs with
total vorticity zero, and as
the temperature is increased across $T_{c}$ from below these vortex pairs start to
unbind~\cite{kosterlitz}. It means that the dominant characteristic physical
features in a region close to the KT transition are associated with vortex pair
fluctuations. One of the physical quantities which contain the information
about the feature of vortex dynamics is the frequency $\omega$ dependent
complex conductivity $\sigma(\omega)$ of the
sample~\cite{minnhagen:rev,rogers}.  In the presence of 2D fluctuation effects
$\sigma(\omega)$ can be expressed as $\sigma(\omega) = -
\rho_0(T)/i\omega\epsilon(\omega)$, where $1/\epsilon(\omega)$ is the dynamic
dielectric function which describes the effect of pair motion, and $\rho_0(T)$
is the bare superfluid density~\cite{minnhagen:rev,amber}. The measurements of
the superconducting transition by means of $\sigma(\omega)$ in a typical
experiment for a fixed frequency should show a single  peak in
$\mbox{Re}[\omega\sigma(\omega)]$ at a frequency dependent
temperature $T_\omega$. This peak represents dissipation losses while a sharp 
decrease in $\mbox{Im}[-\omega\sigma(\omega)]$ at the same temperature is a
consequence of the loss of a superfluid response. Indeed, such behaviors of
$\sigma(\omega)$ have been confirmed in many experiments on 2D superconductors,
as well as on high-$T_{c}$ superconductors~\cite{rogers,flux}.

\begin{figure}[t]
\centering{\resizebox*{!}{5.2cm}{\includegraphics{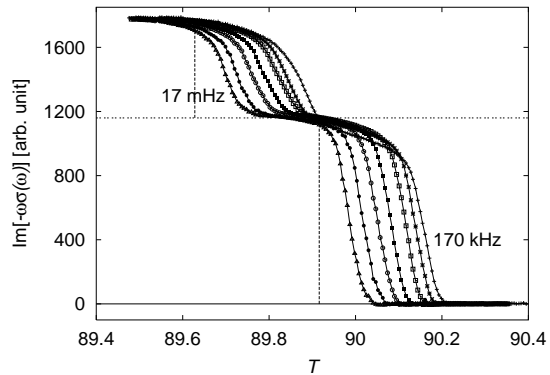}}}
\vskip -0.8cm
\caption{Experimentally measured $\mbox{Im}[-\omega\sigma(\omega)]$  vs $T$  
 for $\omega = 17 \hspace{0.1cm}\mbox{mHz}, 170 \hspace{0.1cm}\mbox{mHz},
1.7 \hspace{0.1cm} \mbox{Hz}, 17 \hspace{0.1cm}\mbox{Hz},
170 \hspace{0.1cm}\mbox{Hz}, 1.7 \hspace{0.1cm}\mbox{kHz},$
$170 \hspace{0.1cm}\mbox{kHz}$ (from the left to the right).
Horizontal lines shows respective
zero levels for transitions and vertical lines indicates  the heights of the
rapid drop of ${\rm Im}[-\omega\sigma(\omega)]$ which gives a rough measure of
the superconducting electrons involved.  
The data are taken from Ref.~\protect\cite{festin}.
}
\label{fig:exp:Im}
\end{figure}

\begin{figure}
\centering{\resizebox*{!}{5.2cm}{\includegraphics{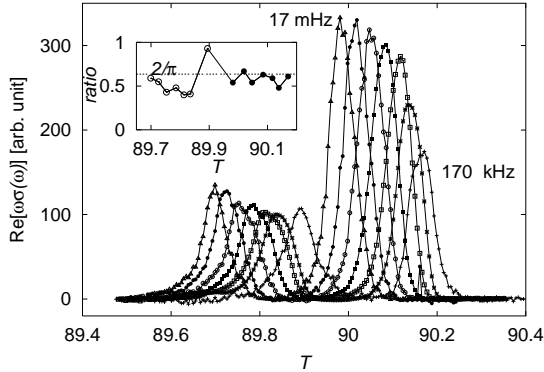}}}
\vskip -0.8cm
\caption{$\mbox{Re}[\omega\sigma(\omega)]$ vs $T$ measured for a thin YBCO film at
 the same frequencies as $\mbox{Im}[-\omega\sigma(\omega)]$
in Fig.1. The inset illustrates the peak ratio defined as
the ratio $\mbox{Re}[\omega\sigma(\omega)]/\mbox{Im}[-\omega\sigma(\omega)]$
taken at the dissipation peak maxima for a given $\omega$.
The data are taken from Ref.~\protect\cite{festin}.}
\label{fig:exp:Re}
\end{figure}

The present investigation is inspired by the recent experimental results
obtained by Festin {\it et al.}~\cite{festin} for a $1500$ {\AA} thin YBCO
film: A very striking double peak structure in
$\mbox{Re}[\omega\sigma(\omega)]$ is found; two rapid drops of
$\mbox{Im}[-\omega\sigma(\omega)]$ at different $T$ are observed.
The data by Festin {\it et al.} are
reproduced in Figs.~\ref{fig:exp:Im} and \ref{fig:exp:Re}. One
possible explanation for the double peak is that the epitaxial grown YBCO film
is split into an upper and lower part due to a slightly different oxygen
contents~\cite{private}.  From this perspective the sample would consist of two
weakly coupled parallel superconducting parts. 

In order to investigate this scenario we use the two weakly coupled 
2D $XY$ models, the Hamiltonian of which is written as
\begin{eqnarray}
H &=& H_1 + H_2 + H_{\rm int}, \nonumber \\
H_1 &\equiv& -J_1 \sum_{\langle ij\rangle} \cos(\theta_i^{(1)} - \theta_j^{(1)}),\nonumber \\
H_2 &\equiv& -J_2 \sum_{\langle ij\rangle} \cos(\theta_i^{(2)} - \theta_j^{(2)}), \nonumber\\
H_{\rm int} &\equiv& -J_\perp \sum_i \cos(\theta_i^{(2)} - \theta_i^{(1)}),\nonumber
\end{eqnarray}
where $H_1$ and $H_2$ are the usual 2D $XY$ Hamiltonians with the coupling
strengths $J_1$ and $J_2$ for the first (lower) and the second (upper) planes, respectively,
the summation $\sum_{\langle ij\rangle}$ is over all nearest neighbor pairs in each
plane, and $H_{\rm int}$ with the coupling strength $J_\perp$ describes the
coupling between the planes. To study the dynamics of the system, 
we use the equations of motion of the standard resistively-shunted junction 
(RSJ) dynamics subject to the periodic boundary condition
and integrate the equations of motion using the second-order
algorithm with the time step $\Delta t = 0.05$. We also apply the fast
Fourier transformation method to speed up the calculations
(see e.g., Ref.~\cite{beom:big} for details).

The 2D $XY$ model on the square lattice undergoes a KT transition at $T_c
\approx 0.89J$~\cite{olsson}, where $J$ is the Josephson coupling strength.
Accordingly, two planes with different coupling constants then 
undergo two separate phase transitions at the different temperature
$T_c\approx 0.89 J_1$ and $0.89 J_2$ when the interplane
coupling vanishes.
The main output from the simulation are the dynamic dielectric function
$1/\epsilon(\omega)$, and the helicity modulus $\gamma$, which basically measures
the stiffness of the system to a twist in the phase of the order parameter and is
proportional to the renormalized superfluid density $\rho =
\rho_0/\epsilon(0)$~\cite{minnhagen:rev}. From a knowledge of
$1/\epsilon(\omega)$ and $\rho_0$ it is straightforward to analyze the behavior
of the conductivity $\sigma(\omega)$.  

%Figure 1
%Figure 2

In order to get realistic parameters for the two weakly coupled 2D $XY$ models
in connection with the experimental result by Festin {\it et al.} one would like to
have some reasonable estimates of the coupling constants
$J_1$, $J_2$, and $J_{\perp}$. We estimate $J_1$ and $J_2$ from the experimental data
reproduced in Fig.~\ref{fig:exp:Im}. In order to do it  we use the relation
$\lim_{\omega\rightarrow 0}\mbox{Im}[-\omega\sigma(\omega)]
= \rho_{0}/\epsilon(0)$ and note that $1/\epsilon\approx 1$ just below
the transition and is 0 just above. Thus $\rho_0\propto J$~\cite{kosterlitz}
may be roughly estimated by the two heights of the rapid drops in
Fig.~\ref{fig:exp:Im} for the curve corresponding to the smallest
frequency. From this we get the ratio $J_2/J_1\approx 0.5$ (ratio between the
dashed vertical lines in Fig.~\ref{fig:exp:Im}). The corresponding curves
for ${\rm Re}[\omega\sigma(\omega)]$ are shown in Fig.~\ref{fig:exp:Re}, where
the characteristic double peaks at different temperatures are clearly exhibited.
$J_\perp$ can be estimated from the knowledge of the anisotropy parameter $\Gamma$
defined as $\Gamma \equiv \sqrt{J_1/J_\perp}$ and found to be equal to $7$ for YBCO
\cite{shilling,nguyen}.
%Figure 3
\begin{figure}[t]
\centering{\resizebox*{!}{5.2cm}{\includegraphics{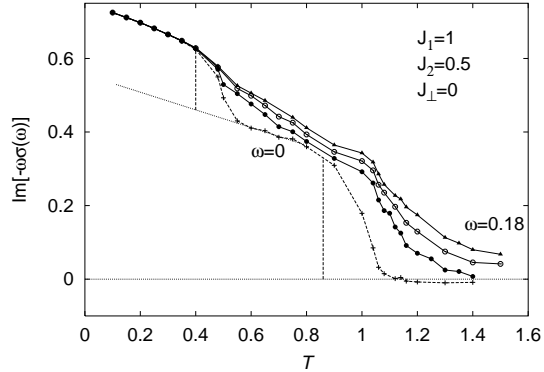}}}
\vskip -0.8cm
\caption{$\mbox{Im}[-\omega\sigma(\omega)]$  vs $T$  for two weakly coupled  2D $XY$
models with system size $L=64$ at  $\omega = 0, 0.06, 0.12, 0.18$ (from the left to the right).
The coupling constants are chosen to be $J_1 = 1$, $J_2= 0.5$, and $J_{\perp} = 0$.}
\label{fig:J0J0.50freqIm}
\end{figure}

%Figure 4
\begin{figure}
\centering{\resizebox*{!}{5.2cm}{\includegraphics{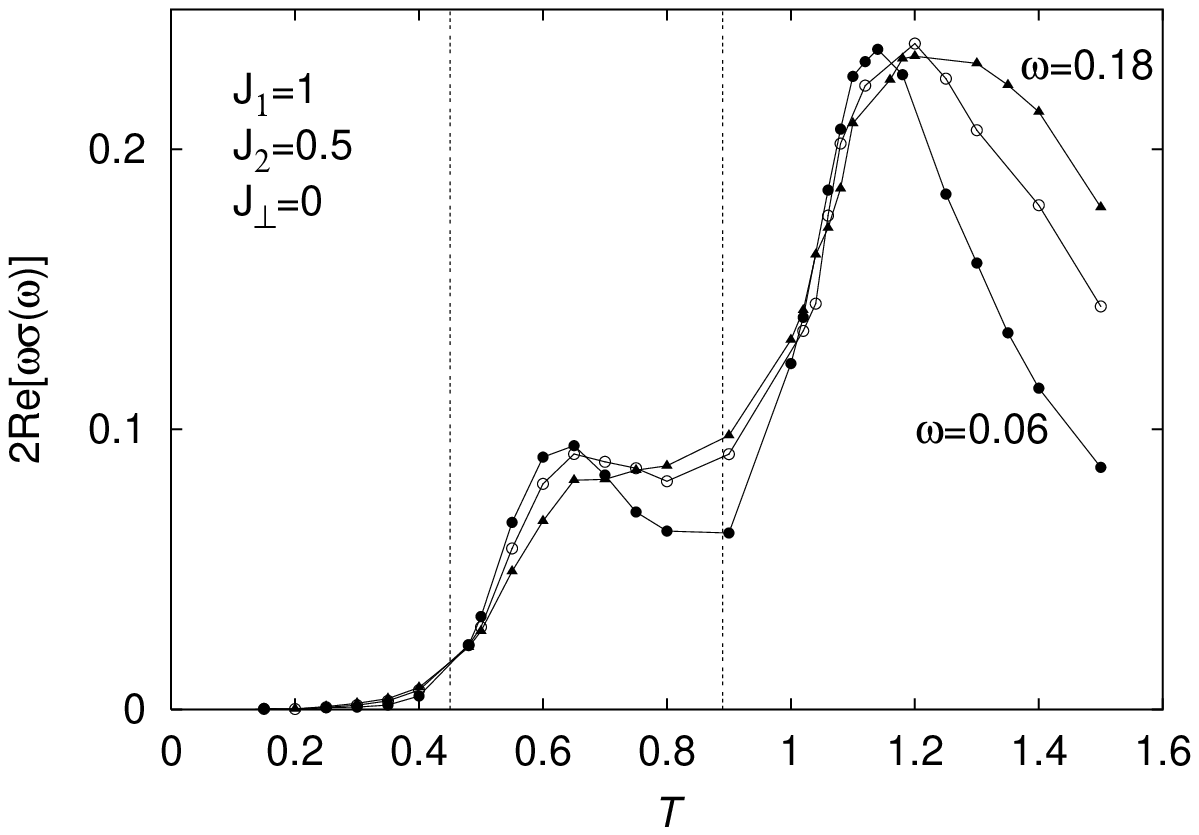}}}
\vskip -0.8cm
\caption{$\mbox{Re}[\omega\sigma(\omega)]$ as a function of temperature
 for different $\omega$: the lines with filled circles, empty circles and triangles
 correspond to the frequency $0.06, 0.12$, and $0.18$, respectively. Vertical lines show
 $T_c$ for two decoupled 2D $XY$ models with  coupling constants $J_1=1$ and $J_2=0.5$:
  $T_{c1}\approx 0.89$ and $T_{c1}\approx 0.45$.}
\label{fig:J0J0.50freq:dif}
\end{figure}

First we present the result for our model without coupling between the two planes.
$J_\perp = 0$ corresponds to the case of no supercurrent flowing between the planes.
Figures~\ref{fig:J0J0.50freqIm} and \ref{fig:J0J0.50freq:dif} show the behavior
of $\mbox{Re}[\omega\sigma(\omega)]$ and $\mbox{Im}[-\omega\sigma(\omega)]$
obtained from the simulations with $J_1=1$, $J_2=0.5$ and $J_\perp = 0$.
The similarities to the experimental
results in Figs.~\ref{fig:exp:Im} and \ref{fig:exp:Re} are striking:
$\mbox{Re}[\omega\sigma(\omega)]$ in Fig.~\ref{fig:J0J0.50freq:dif} again 
displays two distinct peaks  at different 
temperatures while $\mbox{Im}[-\omega\sigma(\omega)]$ in Fig.~\ref{fig:J0J0.50freqIm}
has two regions with increased drops as a function of $T$. Dashed line in 
Fig.~\ref{fig:J0J0.50freqIm} represents
the behavior of $\gamma$ or equally the behavior of
$\mbox{Im}[-\omega\sigma(\omega)]$ in the limit $\omega \rightarrow 0$ which
shows one drop at $T_{c1} \approx 0.89$ and another at $T_{c2} \approx 0.45$. 
These similarities between experiments and our simulations  for the 
coupled 2D $XY$ models are further substantiated when one compares
the frequency dependences:  As the frequency is increased both
$T_{c1}$ and $T_{c2}$ increase as is  reflected in the positions of peaks in
Figs.~\ref{fig:exp:Re} and \ref{fig:J0J0.50freq:dif}. 
%At the same time the heights of the dissipation
%peaks decreases with increasing $\omega$ in both the experimental and
%simulation data.  
The soundness of the  method of estimation of the coupling
constants $J_1$ and $J_2$ is illustrated in Fig.~\ref{fig:J0J0.50freqIm} where
the helicity modulus is represented by the dashed line. Following the
suggestion that $J_2/J_1$ can be estimated from the ratio between the dotted
vertical lines we again get $J_2/J_1 \approx 0.5$.  The quantitative
differences between the data in Fig.~\ref{fig:exp:Re} and the simulations
in Fig.~\ref{fig:J0J0.50freq:dif} is according to our interpretation due to the
fact that the simulations cannot reach as low frequencies as the experiments
(more precisely the ratio between the frequency and the microscopic frequency
scale is larger in the simulations). A higher frequency broadens and smears the
dissipation peaks.
Since the interplane coupling in experiments presumably does not vanish, we have also
investigated the influence of a small $J_\perp$ in our model
and found that a weak interplane coupling ($0 \leq J_\perp \leq 0.04$) does not
influence the  qualitative features found above for $J_\perp=0$.

%We have also simulated the coupled 2D $XY$ model for a larger $J_1$, 
%i.e. $J_1=1$, $J_2=0.75$, and $J_{\perp}=0.04$. 
%The data presented in Fig.~\ref{fig:J0.75J0.04multi} are for the lowest
%$\omega$ one can achieve in simulations.
%Indeed, as one can see from the inset, $\mbox{Re}[\omega\sigma(\omega)]$ of each of the planes
%displays a peak. These two peaks are shifted in $T$. However, this shift is modest
%and it results in a single peak in $\mbox{Re}[\omega\sigma(\omega)]$
%for the whole sample. Thus, with the choice of
%the coupling constants  $J_1=1$, $J_2=0.75$, and
%$J_{\perp}=0.04$ we cannot reproduce the experimental result presented in
%Fig.~\ref{fig:exp:Re}. Even more, the data for the case
%$\omega=0$ corresponding to $\gamma$ (dashed line in
%Fig.~\ref{fig:J0.75J0.04multi}) suggests a
%single transition because when $J_2/J_1$ becomes close to unity the
%transition in the two planes overlap.

The similarity with the experimental data by Festin {\em et al.}
implies that the YBCO film is split parallel to the surface
into two superconducting parts both having transitions
with 2D character but with different $T_c$. Since the coupling between the planes does
not wash away the double peak transition one must infer that the
coupling between the two sheets is very weak.
One possible reason for the  phase separation may lie in a combination of an inhomogeneous surface with a more homogeneous part closer to the substrate. This can lead to different oxygen contents in the two parts, which in turn results in slightly different lattice parameters causing a physical boundary between the two parts~\cite{festin,nakazawa,claus,janod} 

%Figure 5
%\begin{figure}
%\centering{\resizebox*{!}{5.2cm}{\includegraphics{J0.75J0.04multi.eps}}}
%\vskip -0.8cm
%\caption{$\mbox{Re}[\omega\sigma(\omega)]$ and
%$\mbox{Im}[-\omega\sigma(\omega)]$ as a function of temperature for the fixed
%$\omega$ with the set of the coupling constants $J_1=1$, $J_2=0.75$, and
%$J_{\perp} = 0.04$. Dashed line corresponds to $\gamma$. The inset shows three
%different parts that build up $\mbox{Re}[\omega\sigma(\omega)]$ of the sample.
%(Filled and open triangles correspond to the contributions
%from two 2D $XY$ models, the crosses are interplane contribution.
%Summation of these three contributions gives the total $\mbox{Re}[\omega\sigma(\omega)]$,
%denoted by filled circles.) } 
%\label{fig:J0.75J0.04multi}
%\end{figure}

A further consistency check on the scenario in terms of two coupled 2D
superconducting parts comes from the peak ratio defined as the ratio
 $\mbox{Re}[\omega\sigma(\omega)]/\mbox{Im}[-\omega\sigma(\omega)]$
taken at the dissipation peak maxima for a given $\omega$. For a 2D
superconductor this peak ratio should vary between $2/\pi\approx 0.63$
for small $\omega$ to 1 for larger $\omega$~\cite{jonsson}.
The inset in Fig.~\ref{fig:exp:Re} gives the ratios estimated from the
experimental data. The filled circles is for the large-$T$ peaks and these ratios
are consistent with 2D vortex fluctuations close to the transition.
The empty circles corresponds to the small-$T$ peaks where in
accordance with an interpretation in terms of two transitions the zero
level of $\mbox{Im}[-\omega\sigma(\omega)]$
for the low-$T$ transition is estimated by a linear extrapolation of
the large $T$-part towards lower $T$. An almost as good estimate is to
approximate by the plateau between the transitions (horizontal line in
Fig.~\ref{fig:exp:Im}). These peak ratios are also rather consistent with a
2D transition in the limit of small $\omega$ (peak ratios for the lowest-$T$ values).
The small deviation for higher temperatures  may be due to a convolution of the high
and low $T$ parts of the transition.
  
Another question is 
at what thickness $d$ the division takes place. A rough estimate may be
obtained by assuming that both sheets consist of identical material.
In such a case $J_1\propto d_1 n_S$ and $J_2\propto d_2 n_S$ where $
d_{1(2)}$ is the thickness of respective sheets and $n_S$ is the
density of Cooper pairs for the material. Thus $J_2/J_1\approx d_2/d_1$ 
and since we have found that $J_2/J_1\approx 0.5$ from the data, we conclude that the boundary between the two parts should occur somewhere in the middle of
the sample.

In summary we conclude from our numerical simulations of two weakly
coupled $XY$ models that the double peak dissipation structure
observed by Festin {\it et al.} is consistent with the interpretation
that the sample consists of two parts with slightly different
transition temperatures, and that these two parts are separated by a boundary which
mainly runs parallel to the substrate.

%\begin{acknowledgments}
The authors acknowledge \"{O}. Festin and P. Svedlindh for providing their
experimental data.  This work was supported in part by the Swedish Natural Research Council
through Contract No.\ F 5102-659/2001.
%\end{acknowledgments}

\end{document}